# Quantum Encryption in Phase Space for Coherent Optical Communications

Adrian Chan[1], Mostafa Khalil[2], Kh Arif Shahriar[2], David V. Plant[2], *Fellow, IEEE*, Lawrence R. Chen[2], *Senior Member, IEEE*, Randy Kuang[1], *IEEE*

*Abstract*— Optical layer attacks on communication networks are one of the weakest reinforced areas of the network, allowing attackers to overcome security when proper safeguards are not put into place. Here, we present our solution–Quantum Encryption in Phase Space (QEPS)–, a physical layer encryption method to secure data over the optical fiber, based on our novel round-trip Coherent-based Two-Field Quantum Key Distribution (CTF-QKD) scheme. We perform a theoretical study through simulation and provide an experimental demonstration. The same encryption is used for QEPS as CTF-QKD but achieved through a pre-shared key and one-directional transmission design. QEPS is uniquely different from traditional technology where encryption is performed at the optical domain with coherent states by applying a quantum phase-shifting operator. The pre-shared secret is used to seed a deterministic random number generator and control the phase modulator at the transmitter for encryption and at the receiver for decryption. Using commercially available simulation software, we study two preventative measures for different modulation formats which will prevent an eavesdropper from obtaining any data. QEPS demonstrates that it is secure against tapping attacks when attackers have no information of the phase modulator and pre-shared key. Finally, an experiment with commercial components demonstrates QEPS system integrability.

## I. Introduction

WITH increasing demand for faster, more affordable, and smaller form factor solutions in optical communications, the security of the optical network becomes essential in protecting the immense amount of information that is transmitted. Currently, optical networks are mainly secured by protocols at the second layer of the OSI model and above, relying on a software-based solution to secure communication [1]. However, security threats at the physical and optical layer cannot be ignored as eavesdroppers can have unfettered access to the optical layer and potentially compromise data transmitted. With the major development in recent years with quantum computers, ciphertexts are potentially vulnerable and can be cracked in significantly shorter periods of time. By utilizing an optical layer encryption, security can be increased and resist attacks at these layers. Applying an optical layer encryption in the system will allow for low latency, protocol agnostic, enhanced security and transparent end-to-end communication can be achieved.

Various methods have been proposed and implemented for enhanced security in the physical layer of the system including optical code division multiple access (O-CDMA) [2], optical chaos signal generation [3], optical steganography [4] and XOR encryption [5]. These methods have shown vulnerabilities against attacks such as optical chaos signal generation and optical steganography being vulnerable to time-delay identification and post-processing statistical methods. The security performance of O-CDMA has been investigated thoroughly but remain an open issue and are highly dependent on system design and implementation [6]. Similarly, alternative secure communication systems have been presented for key distribution using quantum mechanics such as Quantum Key Distribution (QKD) which is based on the theoretically secure BB84 protocol [7]. One common layout is continuous-variable QKD (CV-QKD) which is based on using an amplitude such that the on-state is at a level that prevents an eavesdropper, Eve, from discriminating the signal and allowing Alice and Bob to detect an eavesdropper by comparing measurements and the variance of the distribution [8]. Gaussian Modulated Coherent State CV-QKD has demonstrated its capability of reaching a secure key rate of 7.04 Mbps over 25 km of fiber [9]. These systems can be used in parallel with conventional data communications through DWDM multiplexing/demultiplexing techniques to achieve both data transmission and secure key distribution [10]. Another common and more recent layout is twin-field QKD (TF-QKD) [11] which is capable of overcoming the PLOB-repeaterless bound [12]. Recently, this system has demonstrated its capability of reaching over 800 km at a secret key rate of 0.014 bps by utilizing a four-phase twin-field protocol and high-quality set-up [13]. These systems, although scientifically secured with its proven security against general attacks and information-theoretic security, are not recommended by major entities such as the National Security Agency (NSA) [14]. This is because security models are unable to encompass all features of a real-world component in preparation and detection and can only provide a guidance where each specific set-up must undergo a thorough study [15]. More work must be accomplished in this field before it can be adapted for commercial use. Nevertheless, QKD, on its own, is only capable of distributing keys used for digital data

1. Quantropi Inc., Ottawa, ON K1Z8P9 Canada. (email: adrian.chan@quantropi.com; randy.kuang@quantropi.com).
2. Photonics System Group, Department of Electrical and Computer Engineering, McGill University, Montreal, QC H3A0G4, Canada (email: mostafa.khalil2@mail.mcgill.ca; kh.shahriar@mail.mcgill.ca; david.plant@mcgill.ca; lawrence.chen@mcgill.ca).

encryption and its current limitations prevent it from being used for direct quantum encryption for high-speed coherent optical transmission [10, 16].

Alternative security measures are currently put into place, but mainly focus on algorithmic encryption. The most common of these encryptions is AES with an encryption through classical public key algorithm such as RSA and Diffie-Hellman. These classical algorithms have been proven to be vulnerable to quantum computing. Quantum encryption applied directly in the optical domain is a unique method which can create a system that maintains confidentiality and integrity, as well as minimize latency.

Here, we present the design, simulation, and preliminary experimental results of a new optical layer security design, Quantum Encryption in Phase Space (QEPS). QEPS exploits quantum phase-shifting operators to quantum mechanically encrypt the optical signals. QEPS is a novel symmetric encryption scheme which is extended from the asymmetric encryption presented by Kuang and Bettenburg in 2020 [17], also known as Coherent-based Two-Field QKD (CTF-QKD), which was developed as an alternative to the QKD protocol by utilizing a public key envelope. QEPS was developed to be used after key distribution between Alice and Bob using techniques such as CTF-QKD. Therefore, the asymmetric encryption will be used to establish the shared key while the symmetric encryption scheme will be used for data encryption after the secret key has been shared between the two users. With little to no difference in equipment required between these two protocols, the asymmetric encryption can easily be switched to the symmetric encryption. Like CTF-QKD, QEPS generates an envelope based on the pre-shared secret, then performs a standard modulation scheme to encode data at the transmission side. QEPS will then remove the envelope based on the same pre-shared secret before coherent detection or through digital signal processing (DSP) after detection. This allows the transmission to be performed in one-direction, from Alice to Bob, to maintain the confidentiality and integrity of the system. Confidentiality and integrity are maintained by leveraging a deterministic Random Number Generator (RNG) driven phase encoding to generate an envelope at Alice transmission (Tx) to encrypt the data that she will send to Bob receiver (Rx). Generating an envelope will provide quantum security in an existing infrastructure while having minimal impact on the performance of the optical communication system. The security of this system follows the same quantum encryption that has been described in detail in [18] and is based on leveraging Glauber states and the number-phase uncertainty principle, $\Delta n \Delta \varphi \geq 1/2$, where an attacker will obtain nondeterministic results from an invisible tap while Alice and Bob are able to operate deterministically. This contrasts with QKD where Alice and Bob are not able to perform normal operation while Bob is able to detect attacks. An analysis will be performed, and the performance of the system will be measured and compared with and without an eavesdropper. OptiSystem 18.0 and MATLAB simulation platforms are used to demonstrate the system performance.

## II. QUANTUM ENCRYPTION IN PHASE SPACE

Quantum encryption in QEPS is based on applying quantum operators to Glauber/coherent states. A quantum phase-shifting operator is applied to the coherent state at the transmitter and is shown as,

$$\hat{U}(\varphi(t))|\alpha\rangle = |\alpha e^{i\varphi(t)}\rangle \quad (1)$$

where $\hat{U}(\varphi(t))$ is the phase-shifting operator driven by an RNG seeded with a pre-shared key.

Leveraging the fact that phase-shifting operators are unitary and reversable, the conjugate transpose can be applied to reverse the phase shifted operation at the receiver by using the same PRNG seeded with the pre-shared key between both the transmitter and receiver operator,

$$\hat{U}(\varphi(t))^\dagger \hat{U}(\varphi(t))|\alpha\rangle = |\alpha\rangle \quad (2)$$

By applying the operator's conjugate transpose, the identity operation is obtained for $\hat{U}(\varphi(t))^\dagger \hat{U}(\varphi(t)) = \hat{I}$ since $\hat{U}(\varphi(t))$ is a unitary operator. Therefore, the original coherent state can be recovered. In the aspect of our technology, the phase-shifting operator and conjugate transpose operator can be applied by a phase modulator or PM before coherent detection or after coherent detection in DSP at the receiver.

The basis of QEPS technology can be related to the operation of CV-QKD. CV-QKD operates by the quadrature of the electric field in the optical phase space by transmitting coherent states from Bob to Alice by randomly selecting between the "off" ($|0\rangle$) or "on" state ($|1\rangle$). The random selection is typically achieved by modulating the phase to impose the $|0\rangle$ with 0 degrees or $|1\rangle$ state with 180 degrees. The amplitude is tuned to a level where the probability distributions of both states overlap. The variance from the measurement will be used to determine whether tampering occurs. In contrast to CV-QKD where the global reference phase is static, QEPS leverages the phase space where the reference space of the in-phase and quadrature operator of the coherent state is manipulated with the phase-shifting operator, making the global reference space dynamic. This dynamically changes in time through phase modulation which is driven by a RNG and decryption can easily be accomplished through a pre-shared key. That is, for QEPS, the "static" global phase can only be established between the trusted transmitter and receiver with the pre-shared key. In detection, if the expected BER increases then it can be determined that an eavesdropper was present. Furthermore, even with an "invisible" tap, the quantum encryption will result in the eavesdropper obtaining random data with a high BER around 0.5.

This quantum encryption in phase space will secure the optical line by applying different phase shifts to the coherent state preventing an eavesdropper from obtaining any information of the coherent state, creating a noncoherent channel. This is achieved when an eavesdropper taps the fiber. The tapped signal that an eavesdropper will have to decrypt is $|\alpha e^{i\varphi(t)}\rangle$, where the coherent state is masked by the quantum phase shifted operation. With a time-varying phase shift, the coherent state will remain random, masked, and secure against an attack. Furthermore, when an eavesdropper taps a signal, based on the number-phase uncertainty principle, the phase will be changed. This phenomenon occurs because average number of photons ($\Delta n_p$) decreases, resulting in a larger phase uncertainty ($\Delta \varphi_p$). The $\Delta \varphi_p$ will play a major role for security in the higher modulation schemes such as 128-QAM. Additionally, in practice, there will be differences between the receiver's local oscillator (LO) and the eavesdropper's LO, which will increase measurement error. Therefore, it is more exact to describe the tapped signal that an eavesdropper will have to decrypt as $|\alpha e^{i(\varphi(t)+\Delta\varphi_p+\Delta\varphi_{LO})}\rangle$.

It is also important to touch on the generator that will determine $\varphi(t)$. $\varphi(t)$ is driven by a RNG component such as the deterministic pseudo-Quantum RNG (pQRNG). Other RNGs such as a genuine QRNG can be used, however require a large set of random numbers pre-shared between Alice and Bob and used repeatedly between a synchronized QEPS encryption and decryption. Another alternative to apply a genuine QRNG is to send the generated values over another channel to drive the encryption, similar to chaotic phase scrambling [20]. Deterministic RNGs are used in this study to simplify the synchronization between encryption and decryption. The deterministic RNG unit must have good randomness to prevent correlation of future values and a long secret to increase the difficulty of decoding. One class of RNGs that fit these requirements is pQRNG such as the one described in [21]. The pQRNs in [21] can be generated by supplying random numbers to select specific permutation matrices in the quantum permutation pad. Utilizing the novel quantum permutation pads, the pQRNG can hold over 100,000 bits of entropy with 64 8-bit permutation matrices through the pre-shared secret to deterministically drive the phase-shifting operator, providing added security to the system. The entropy of this pQRNG can quickly be scaled to increase the security by increasing the bits of the system and number of permutation matrices. The pQRNG is seeded with a pre-shared secret of up to 16 kB supplied by a telco operator. This method will prevent an eavesdropper from decrypting the phase randomization both physically and digitally and maintain the confidentiality of the data over the fiber optical layer.

Finally, QEPS's security can be described through the calculation of mutual information like a noncoherent channel where the phase information is unable to be transmitted over the channel. The description below will extend the work performed in [22] with the addition of our QEPS implementation. The QEPS system will be assumed to follow a noncoherent channel as the phases will be completely randomized through a uniformly distributed deterministic RNG. Noncoherent channels can be used to model QEPS as they are AWGN channels which have introduced random phase rotations [23]. The random phase rotations in the QEPS case are a result of the quantum phase-shifting operators applied onto the optical signal. Randomization is achieved through the phase modulator (PM) driven by an RNG. This noncoherent channel will provide a lower bound of information leakage that Eve will be able to obtain in the ideal situation. Indeed, a complete analysis will be performed in the future to identify realistic cases with limited number of phase slices, maximum phase shifts applied, and limiting the tapping power. Without the quantum encryption from QEPS, data travels through a coherent channel containing information from both the phase and amplitude depending on the modulation format. Based on the SNR, a malicious party can obtain information and will be explained from the following model. The channel will have a complex-valued input,

$$X = X_\| \cdot e^{jX_\sphericalangle}, \quad X_\| \in [0,\infty), X_\sphericalangle \in [-\pi,\pi) \quad (3)$$

and a complex-valued output,

$$Y = Y_\| \cdot e^{jY_\sphericalangle}, \quad Y_\| \in [0,\infty), Y_\sphericalangle \in [-\pi,\pi) \quad (4)$$

For a partially coherent the continuous-time form can be described by,

$$Y(t) = X(t) \cdot e^{j\Theta(t)} \cdot e^{j\varphi(t)} + N(t) \quad (5)$$

where $\Theta(t)$ is the phase noise process, $\varphi(t)$ is the phase space randomization and $N(t)$ is the complex-valued additive white gaussian noise (AWGN) process with a variance of $2\sigma_n^2$. An ideal interleaver and de-interleaver can convert (3) into the discrete-time form following the form,

$$\begin{aligned} Y_i &= X_i \cdot e^{j\Theta_i} \cdot e^{j\varphi_i} + N_i \\ &= (X_i + N_i) \cdot e^{j\Theta_i} \cdot e^{j\varphi_i} \\ &= X_i \cdot e^{j\Theta_i + \varphi_i} + N_i' \end{aligned} \quad (6)$$

where $N_i' \sim N_\mathbb{C}(0, 2\sigma_n^2)$.

Polar decomposition of mutual information for an AWGN channel with Gaussian input, phase noise and quantum encryption can be calculated through the equations presented in [22]. The mutual information $I(X;Y)$ is described as,

$$I(X;Y) = I(X_\|;Y_\|) + I(X_\sphericalangle;Y_\sphericalangle|X_\|) + I(X_\|;Y_\sphericalangle|Y_\|) + I(X_\sphericalangle;Y_\||X_\|,Y_\sphericalangle) \quad (7)$$

where the terms on the right side of the equation from left to right represents the Amplitude term, Phase term, Mixed term I, and Mixed term II. Mixed term I is the amount of information about the input amplitude that can be drawn from the output phase given the output amplitude. Mixed term II is the amount of information about the input phase that can be observed from the output amplitude given the input amplitude and output phase. This polar decomposition represents the information that is sent in each component and can be related to the total amount of information that attainable by the receiving party or a malicious party in a coherent channel. The polar decomposition of the mutual information was plotted in Fig. 1(a) by Goebel *et al.* for 16-QAM [22].

QEPS can easily be integrated for standard data modulation formats. As the phase is completely randomized through the pseudo-random selection of phases, and assuming that an infinite number of phases between [-$\pi$, $\pi$) can be chosen and is uniformly distributed, the output phase $y_\sphericalangle$ will contain no information resulting in the Phase term and Mixed term I to equal zero. Since $y_\sphericalangle$ carries no information, the Phase term will also have no mutual information available resulting in the Phase term to be zero. Mixed term I tends toward zero because it becomes a continuous concentric ring with an infinite number of phases. Mixed term II will also be assumed to equal zero due to $p(y_\||x_\|,y_\sphericalangle) = p(y_\||x,y_\sphericalangle)$. Therefore, the mutual information only contains the Amplitude term, $I(X;Y) = I(X_\|;Y_\|)$. Indeed, this scenario is ideal and in practice, a finite number of phase levels would be chosen where the Phase term would have some value as long as the variance is small and the Mixed term II would have a negligible small value. This will be the mutual information that travels through the fiber and also the maximum information that Eve can obtain if she taps 100% of the power. The ideal polar decomposition of mutual information for Eve with QEPS applied to 16-QAM is shown in Fig. 1(b). As expected, the maximum and ideal mutual information that Eve can obtain from a noncoherent channel is significantly reduced compared to Fig. 1(a) and in a realistic scenario where Eve would only tap a small amount of power, the Amplitude term that she would obtain would be even smaller. On the other hand, Bob can convert the noncoherent channel back to a coherent channel by applying the decryption using the pre-shared secret and obtaining the modulation output phase $y_∡$. Doing so will allow Bob to recover the Phase term, Mixed term I, and Mixed term II. Therefore, his mutual information will consist of all the mutual information terms shown in (7) and in the ideal situation will be the same as Fig. 1(a) once decryption is performed. These findings can recommend that QEPS with the phase shift operator is secure for PSK modulation formats where no amplitude information would be present resulting in a mutual information approaching zero in the ideal case.

To summarize, this initial mutual information study with the idea assumption and cases have provided a lower bound and guidance in the leakage of information with the conclusions listed:
1. Bob is able to recover all mutual information terms through decryption with knowledge of the pre-shared key.
2. PSK modulation formats provide the lowest amount of information leakage as the phase slices increase.
3. PAM formats provide the least security as the amplitude term remains through encryption.
4. QAM is a mixture of both amplitude and phase modulation which results some information leakage, mainly due to the amplitude modulation.
5. Eve still obtains cipherbits and is required to know the pre-shared key to obtain all information.

A more comprehensive study will be performed in the future to accurately quantify the security the QEPS in a standalone paper. Nevertheless, the results shown in this mutual information description of QEPS operating over the optical fiber demonstrates that a malicious party will only be able to obtain a minimal amount of information for QAM modulation formats and negligible amounts of information for PSK modulation formats.

## III. System Implementation

The QEPS system schematic is shown in Fig. 2. There are two main sections: the transmitter, where phase encryption and data encoding occur, and receiver, where phase decryption and data decoding occur. The first section, Alice Tx, generates a coherent light which is quantum encrypted by a PM and driven by a deterministic RNG seeded with the pre-shared secret. This PM acts as a quantum phase-shifting operator which randomizes the phase of each coherent state. A pre-shared key is required to allow both users to encrypt and decrypt their data. This requirement may offer telco operators the advantage to control their data security over the infrastructure layer and avoid any possible security backdoor set by optical transceivers.

Alice's data will then be encoded into the phase randomized coherent light through an IQ-MZM system using the modulation of their choice. Alice's quantum encrypted data or optical cipher signal is then sent to Bob. Bob will receive Alice's optical cipher signal and perform coherent detection. Quantum decryption will be performed in DSP where the pre-shared key will be used to remove Alice's phase-shifting operator's effect. This step is performed by applying the conjugate transpose of the initial phase shifted gate digitally to the detected signal. Finally, typical DSP algorithms can be performed to obtain Alice's encoded data. In essence, the major difference between the system presented in Fig. 2 and a conventional coherent optical communication system is the addition of an optical layer quantum encryption at Alice's transmitter and an additional quantum decryption step at Bob DSP.

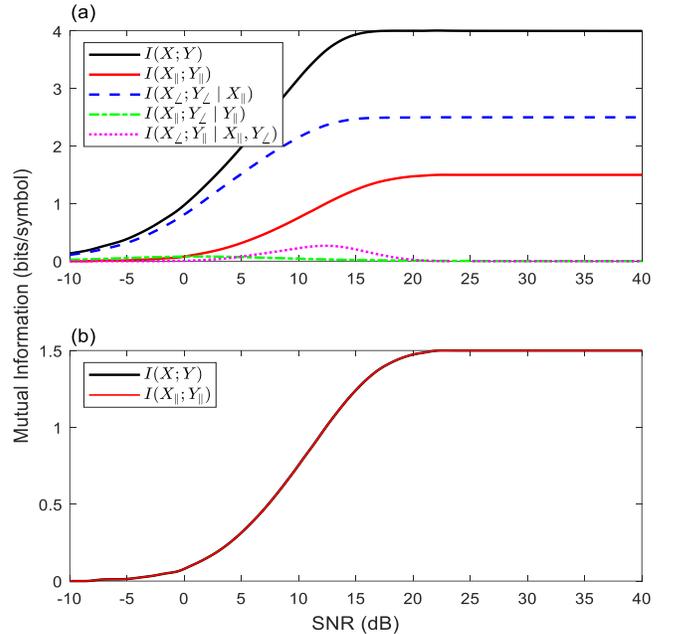

Fig. 1. Polar decomposition of mutual information for (a) 16-QAM [21] and (b) 16-QAM with QEPS applied.

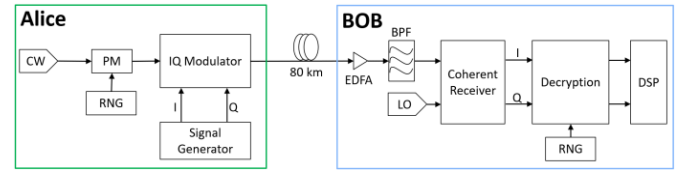

Fig. 2. Schematic diagram of the proposed QEPS system. CW: continuous wave laser, BPF: band-pass filter, I: in-phase signal, Q: quadrature signal.

The communication between Alice to Bob will be described in detail. Firstly, Alice creates a coherent state, $|α_b⟩ = |\sqrt{μ}⟩$, where $μ$ is the signal intensity. She will then apply phase randomization to the optical signal resulting in $|α_b⟩ = |\sqrt{μ}e^{iφ_b}⟩$ where $φ_b$ is the phase randomization applied through a PM driven by a deterministic RNG. Furthermore, this RNG unit will be used by both Alice and Bob together to generate the same randomized pattern to allow for seamless encryption at Alice and decryption at Bob. The phase randomized state will act as an envelope where Alice will then encode her information using their desired standard modulation format, resulting in the

output coherent state $|\alpha'_b\rangle = \left|\sqrt{\mu_b}e^{i(\varphi_b+\varphi_a)}\right\rangle$, where $\varphi_b$ is the phase randomization and $\varphi_a$ and $\mu_b$ represents the intensity modulated data that Alice sends to Bob. The optical signal will then be sent to Bob, where he will perform coherent detection and obtain an optical power that is incident at the photodetectors given as,

$$P = \mu_b + \nu + 2\sqrt{\mu_b \nu}\cos(\varphi_a + \varphi_b) \quad (3)$$

where $\nu$ is the LO intensity at Bob Rx.

In contrast to [17], in (8), the envelope remains at the detection and will be removed digitally. This can be achieved because Bob and Alice have a pre-shared key to apply the phase-shifting operation digitally. Alice's encoded information will then remain, and Bob can apply typical digital signal processing (DSP) algorithms to obtain Alice's encoded data. Typical DSP algorithms include but are not limited to DC blocking, resampling, QI compensation, dispersion compensation, nonlinear compensation, timing recovery, adaptive equalizer, frequency offset estimation, and carrier phase estimation.

The system schematic of a typical attack on the optical fiber where an eavesdropper taps the QEPS system is shown in Fig. 3. In essence, we will be simulating the most desirable attack by Eve where she is next to Alice and has access to her output port. An eavesdropper Rx would receive a power equation like (8). This received signal will be lower in quality and be weaker in power due to Eve only being able to tap a small portion of the transmitted power. With a lower quality signal, the number-phase uncertainty increases $\Delta\varphi_p$ due to $\Delta n$ being small, masking the correct bit and preventing an eavesdropper from correctly obtaining the data transmitted. Therefore, by operating within a range where the amplitude of the signal is deterministic for Alice and Bob whereas a tapped signal demonstrates high BER will allow the system to remain secure against any eavesdropper.

Single polarization simulations will be performed for simplicity; however, this system is not limited to single polarization but can also be extended to dual polarization. Simulation system layout parameters are listed in TABLE I. The PM component has one parameter, the phase deviation, which sets the maximum phase shift applied to the optical signal. System security is maximized by selecting phases between zero and the phase deviation. Furthermore, coherent state phases were randomized with a MATLAB component. One parameter was used in this component, a period parameter to set the length of bits for which a phase would remain constant. Our algorithm uses a constant period that changes the phases after a pre-defined number of bits, this period can also be randomized to further improve security and prevent Eve from determining Bob's secret key; however, for each modulation format, a minimum period value is required at a specific transmission rate due to the BER. Unless otherwise stated, the phase deviation will be set to 90 deg, the period will be set to 1024, and 5 sets of simulations will be performed per analysis with averages and standard deviations shown. The optical signal's power generated at both transmitter (Alice's CW laser) and receiver (Bob's LO) will be matched. Lastly, we assume Kerckhoff's principle [24] for Eve which states that the eavesdropper has the same receiver system as Bob, has an available tapping port at Alice's output, and all information of the system but not the information related to the randomized phase pattern (period value, RNG seed).

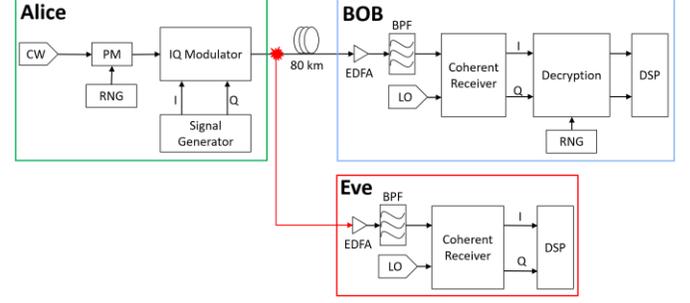

Fig. 3. Schematic diagram of the QEPS system with a tapping attack.

TABLE I
SIMULATION TEST SET PARAMETERS

| | | |
|---|---|---|
| **Layout Parameter** | Sequence length | 65,536 bits |
| | Baudrate | 28 Gbaud |
| | PM period | 1024 |
| **CW Laser & Local Oscillator** | Center wavelength | 1550 nm |
| | Linewidth | 0.1 MHz |
| | Azimuth | 45 deg |
| **IQ Modulator** | Extinction ratio | 20 dB |
| | Switching bias | 3 V |
| | Insertion loss | 5 dB |
| **EDFA** | Forward pump power | 11 mW |
| | Forward pump wavelength | 980 nm |
| | Loss at 1550 nm | 0.1 dB/m |
| | Loss at 980 nm | 0.15 dB/m |
| **Optical Fiber** | Length (1 spool) | 80 km |
| | Attenuation | 0.2 dB/km |
| | Dispersion | 16.75 ps/nm/km |
| | Dispersion slope | 0.075 ps/nm$^2$/km |
| | Differential group delay | 0.2 ps/km |
| | Effective area | 80 μm$^2$ |

## IV. SIMULATION RESULTS

Two main simulations will be performed to determine the performance and system security:

1. CW power analysis
2. PM phase deviation and period parameter analysis

Simulations were performed for phase shift keying (PSK) and quadrature amplitude modulation (QAM) modulation formats (QPSK, 16-PSK, 16-QAM, 32-QAM, 64-QAM, and 128-QAM). For test cases involving an Eve tap, it will be assumed to be "invisible" and ideal where only a maximum of 10% power can be tapped and with no coupling loss. These results demonstrate one random pattern used, however, a more detailed investigation for different random patterns can be found in [18].

Exploiting the observer effect, when Eve taps the optical fiber, a finite $n_p$ will be siphoned off. The variance of the $\Delta\varphi_p$ will increase for the tapped signal for Eve and the remaining data signal for Bob. Of course, the tapped signal comes with bigger $\Delta\varphi_p$ and variance than the signal to Bob. Therefore, we will assume that Eve can only tap a maximum of 10% for all simulations to minimize their possible $\Delta\varphi_p$ at Alice Tx.

## A. CW power analysis

Fig. 5 shows the simulation BER results when varying the CW power of both Alice and Eve. As stated above, a window of operation must be defined, where Eve is unable to determine the correct bits. For QPSK with a period of 1024 and phase deviation of 90 deg, all CW power tested can be used. For 16-PSK with a period of 1024 and phase deviation of 70 deg, a CW power of -5 dBm or greater is required. Similarly for 128-QAM with a period of 1024 and phase deviation of 90 deg, a CW power of 6 dBm or greater is required. Results for other modulation formats exhibited similar performance where more complex formats required working ranges at stronger input power.

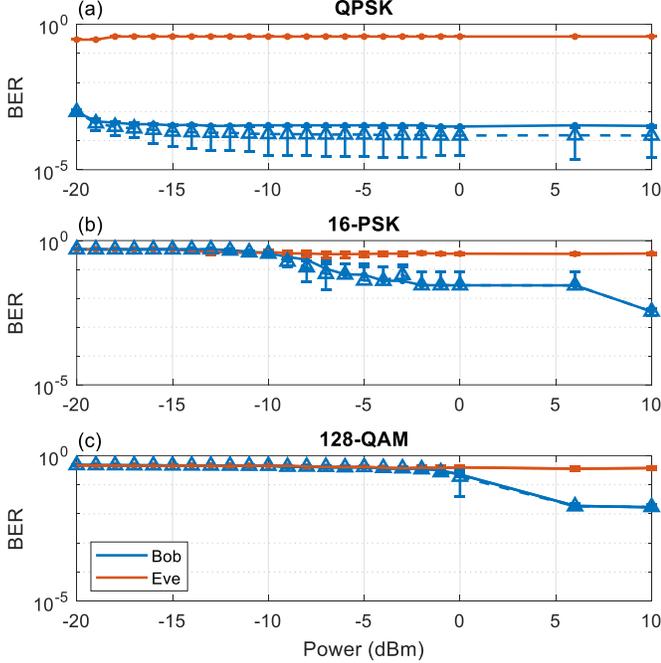

Fig. 5. BER vs. CW laser power simulation results of receiver Bob with an attacker tapping (solid line) and no tapping (dashed line), and attacker Eve for (a) QPSK, (b) 16-PSK and (c) 128-QAM.

In Fig. 4, we demonstrate the security of QEPS with the initial constellation diagrams. In Fig. 4(a)-(b), we simulate a tapping attack with detections at Bob and Eve when no quantum encryption at Alice is applied. Without quantum encryption, the eavesdropper will clearly be able to obtain a relatively low BER value, allowing them to obtain information Alice sends to Bob. This means, a copy of any transmitted data, either encrypted ciphers or plaintexts, through fibers would be obtained by eavesdroppers. That is a fundamental fact for today's optical infrastructure. However, with quantum encryption this capability is disabled for an eavesdropper as shown in Fig. 4(c)-(e). The constellation diagram that Bob receives without Eve's tapping, is clear and each sample can be determined to the correct constellation point with minor errors. With the addition of a tapping attack, the constellations that Bob receives, are relatively good, but with noticeably more error than without a tapping attack. The constellations that Eve receives, are a randomized cluster that results in high indiscernible information with a BER of 0.5. This randomized cluster is a result of the encryption applied where each bit has been randomly shifted, changing the position, and making the constellation diagram for an eavesdropper uninterpretable.

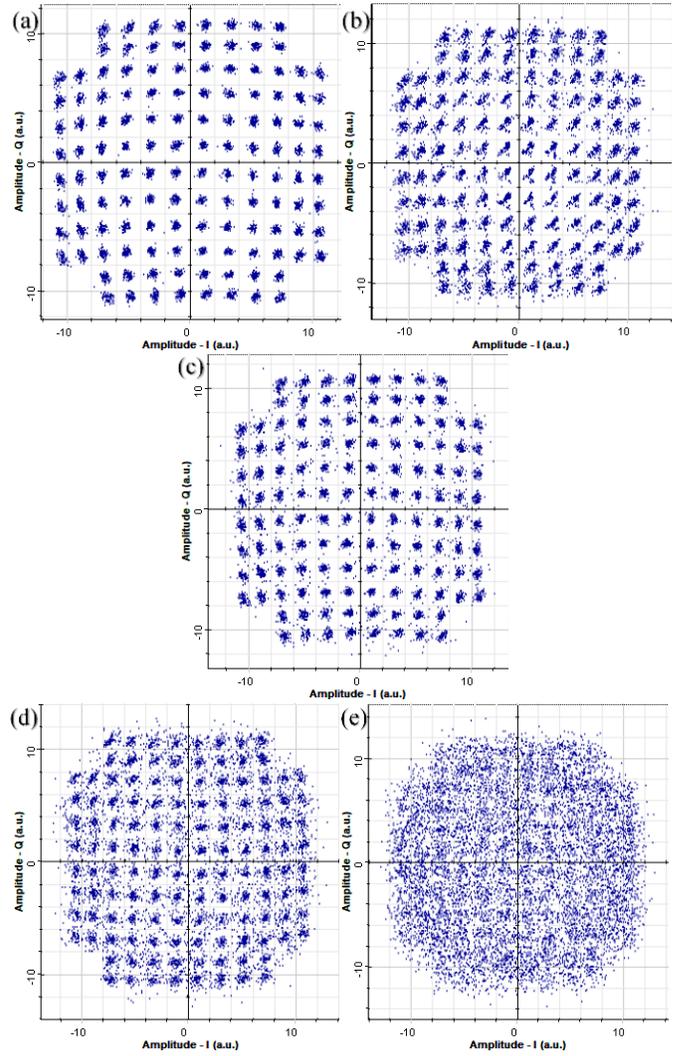

Fig. 4. Constellation diagrams for 128-QAM without QEPS for Bob with no tapping (a), and Eve (b). Constellation diagrams for 128-QAM with QEPS for Bob with no tapping (c), Bob with tapping (d), and Eve (e).

Finally, these results are in line with our experimental results which were presented in [25] for the QPSK case for CTF-QKD. Although the work presented here are for QEPS and not CTF-QKD, we can notice that the quantum encryption applied are similar, but for a different application. This is exactly what QEPS means for optical communication security. This effect can not be seen from other encryption methods such as data encryption in higher layer or physical encryption performed digitally with AES.

## B. PM phase deviation and period parameter analysis

The PM phase deviation and period parameter also play a vital role in determining the security of the system. Phases for every period will be selected between zero and the phase deviation value. It was determined in [18] those different parameters had a large effect on the BER. In this analysis, we will perform similar performance tests as done in [18, 26] in the presence of a tapping attack. Furthermore, we will compare the BER performance with different parameters in Alice's laser, Bob's reference laser, and Eve's reference laser.

QEPS exhibited similar trends to CTF-QKD for both the period and phase deviation parameter [18], as shown in Fig. 6. With an increase in period, Bob's BER was decreased with a trade-off of lower security or less randomness. When the phase deviation was too small, Eve was able to obtain an error-less BER with a phase deviation of 0 deg (no phase encryption applied) and only a small error at 20 deg. A phase deviation of 20 deg resulted in low error because relatively small phase shifts were applied to the optical signal. Even without any quantum de-encryption applied by Eve, with a less complex modulation formats, Eve will be able to obtain a low BER for small phase deviations due to bits remaining in the correct decision boundary after quantum encryption has been applied. At phase deviations larger than 45 deg for 128-QAM, Eve is unable to decode any information and obtains the maximum BER of 0.5. A simple experimental demonstration is shown in Fig. 7 and the equipment used and experimental setup are similar to [25], however this demonstration is configured for the symmetric encryption rather than the asymmetric encryption (CTF-QKD). This highlights the configurability of the set-up allowing seamless transition from a symmetric encryption to asymmetric encryption. This experiment is performed at 6 GBaud for 16-QAM. This experiment was only continued until the HD-FEC limit was reached and values differ from simulation as the experimental set-up layout parameters are not identical. As seen, the experiment follows the same trend as shown in Fig. 6(a). With increasing phase deviation, the BER increases. Furthermore, constellation diagrams of this experiment are shown in Fig. 8 for three phase deviation points for Bob's receiver. To clarify, this experimental demonstration does not contain any eavesdropper and demonstrates the results for an Alice to Bob transmission without any tapping. As expected, with increasing phase deviation of the quantum encryption, the increasing BER of Bob's receiver can be seen with the received data overlapping in constellation points. With increasing phase deviation, larger error occurs due to bigger phase shifts resulting in a larger BER which is seen in Fig. 8(c) where the constellation points start to overlap. It is also important to note that this is a preliminary experimental validation where future work will be performed to verify the conclusions made in this simulation at higher baud rates, more complex modulation formats, and in the presence of Eve. Nevertheless, these results verify the conclusions made for a non tapping scenario.

Lastly, we identify the effect of the laser linewidth on the security of the system. Essentially, we will be determining the effect of using better laser equipment than previously tested and the effect of an eavesdropper having a superior laser than the one that Alice and Bob use in their set-up. OptiSystem is not capable of completely isolating the effects of QEPS and the linewidth. At the low frequency non-linear regime instabilities may occur and may not be encompassed in this set of simulations which include white noise, flicker, random walk noises [27]. These limitations may not be a dominate effect in the present commercial network, however with constant improvements to technology, these effects must be explored. Shown in Fig. 9 are the results for the analysis on different laser linewidth for 128-QAM. With a smaller linewidth, the BER was expected to drop significantly for Bob, however, the laser linewidth was found to not improve Bob's BER. However, for an eavesdropper, because the quality of their tapped signal and their local oscillator is improved, it was found that the security was decreased, and that Eve was able to obtain more correct bits. Nevertheless, improvements to Eve's BER are only apparent at low to no phase deviations.

From these simulation results it was determined that a set of operating quantum encryption parameters are required to ensure the security of the QEPS system. A summary of these optimal operation parameters is shown in TABLE II.

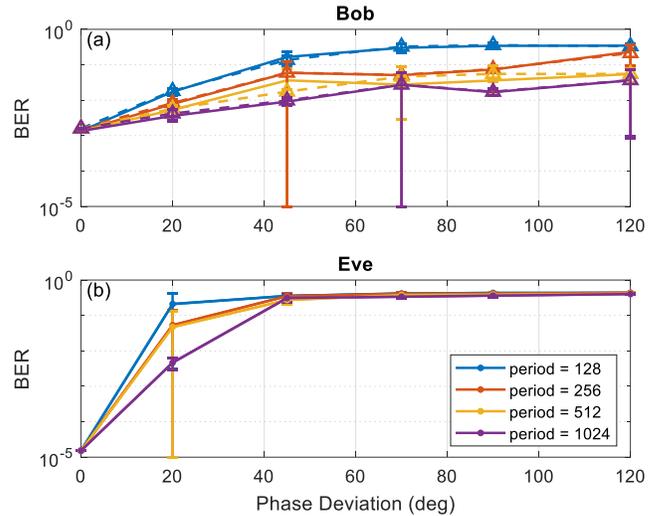

Fig. 6. 128-QAM BER vs. phase deviation simulation results of (a) Bob receiver with an attacker tapping (solid line), and no tapping (dashed line) and (b) Eve for varying periods.

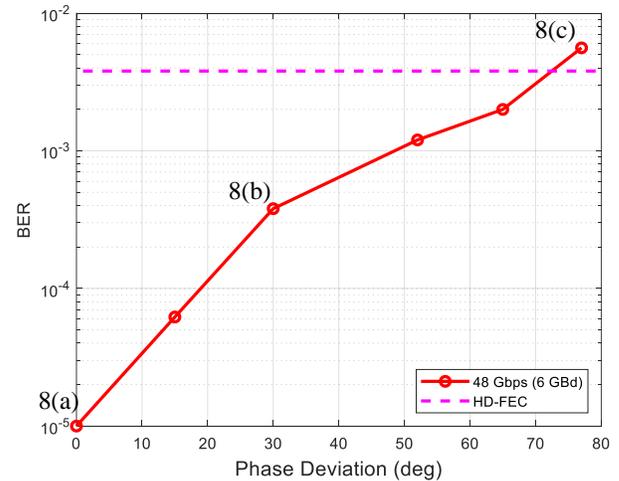

Fig. 7. 16-QAM BER vs. phase deviation experimental results of a Bob receiver with a period of 40. 8(a)-(c) refers to Fig. 8. constellation diagrams.

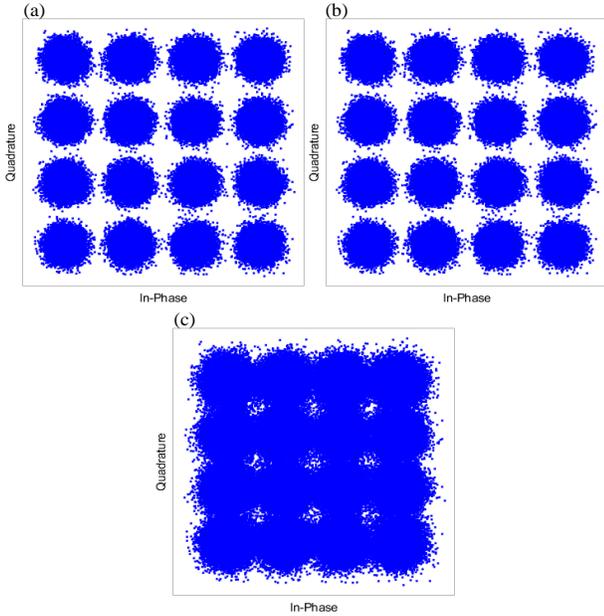

Fig. 8. Experimental constellation diagrams for 16-QAM without QEPS for Bob receiver (a), with QEPS with a phase deviation of 30 deg (b) and 77 deg (c) for Bob receiver.

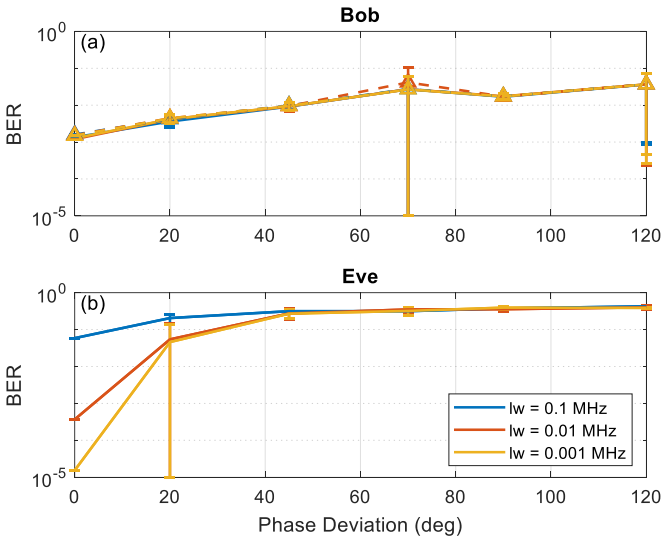

Fig. 9. 128-QAM BER vs. phase deviation simulation results at Bob receiver with an attacker tapping (solid line), and no tapping (dashed line) and (b) Eve for varying linewidths (lw).

TABLE II
OPTIMAL PHASE RANDOMIZATION PARAMETERS

| | | |
|---|---|---|
| **Laser** | Power | Dependent on modulation format |
| | Linewidth | <0.1 MHz |
| **PM** | Phase Deviation | >70 deg |
| | Period | <1024 |

## V. DISCUSSION

In our previous analysis, [18], for CTF-QKD we determined that by controlling the CW power of both Bob Tx and Rx, a window is created where Bob can recover Alice's secret key while preventing Eve from obtaining any transmitted information. Furthermore, we performed an analysis on these results in [26], determining the theoretical and practical security of the system. We have also tested for varying CW power and concluded that Eve is unable to decode for all CW power for all modulation formats, resulting in a BER value of 0.5 at all input power. These results demonstrate that QEPS is compatible with all modulation formats and maintains security by quantum encryption of the encoded data at the transmitter preventing any eavesdropper from obtaining any information. The initial quantum encryption will fully mask Alice's transmission data and can only be decoded with knowledge of the exact phases applied. These results demonstrate the efficacy of applying quantum encryption to the security of the network in preventing any information from being known to an eavesdropper. In contrast to our CTF-QKD results, which described a window of power operation that prevented an eavesdropper from obtaining information, in this system a minimum power is defined to provide security where even at high powers, an eavesdropper is unable to obtain any information. A minimum power is required in QEPS to obtain good BER due to noise from transmission. Additionally, by operating at higher power, lower BER at longer transmission distances can be achieved. It is also important to note that for QPSK modulation, the minimum power required is outside the range of testing shown and was determined to be at a CW power greater than -21 dBm. Although the working range provides a minimum power that the operator must operate at, the BER can also be used to verify if an attacker is presently eavesdropping. In this case, only one specific power can be used, where Bob's BER would increase significantly in the presence of an eavesdropper tapping 10% of their power. For 16-PSK a power of -5 dBm and for 128-QAM a power of 0 dBm. For QPSK, this case would not be possible at any power because there is no significant difference between the BER in the no tapping and tapping case. This is due to the already low operating power that QEPS is capable of operating at for QPSK.

Constellation diagrams were also shown in Fig. 4 to demonstrate the effect that QEPS has on both Bob's and Eve's diagram. Here, the effects of adding quantum encryption to a traditional optical network are compared. Without quantum encryption, it was determined that Eve was capable of obtaining information sent between Alice to Bob assuming that Eve had information of their systems. With quantum encryption, the in-phase and quadrature electrical signals at the Rx are shifted and scrambled, preventing correct detection of bits by Eve. This demonstration is extremely important to the security of QEPS. By changing the phase space of the in-phase and quadrature signals of the data, malicious parties would obtain data that is unidentifiable. The bits of data are mapped to a modulation format which correlates to a constellation point and these constellation points are on the in-phase and quadrature axis. If the phase reference is shifted continuously and without any knowledge of the new phase references, a malicious party would obtain incoherent data. Compared to classical methods, QEPS provides an enhanced or added security as classical methods currently focus on data encryption while QEPS would operate on the optical domain. Therefore, both classical methods and QEPS can be applied together.

We have also shown that the phase deviation and period results demonstrated that selecting large phase deviations can provide greater security in securing Alice's data to Bob. It was also found that Eve could decode the signal up to a maximum of 45 deg. Therefore, it is recommended to use a phase deviation of at least 70 deg for any modulation format tested to prevent an attacker from obtaining any information. An analysis on Eve's phase de-randomization in the DSP was not performed as it has been demonstrated in [19] for various test cases and will be assumed to be similar and comparable in performance when adapted to the QEPS system. Our exploration of transmission impairments on the DSP's ability to distinguish the phase modulation from the encryption process is incomplete and will be performed in the future. This will require more time to test and verify by analyzing factors such as the effect of equalization enhanced phase noise on the DSP's ability to properly perform decryption due to the non-commutative property of the convolution and multiplication. Optimization of DSP algorithms may be required or new algorithms will need to be implemented to compensate for the additional quantum encryption introduced by QEPS. Nevertheless, shown in our experimental results, current DSP algorithms are capable, but not optimized to decode QEPS. It is also important to note that in our simulation results large standard deviations are present within testing. These results are mainly due to one extreme outlier within the test set with a substantial increase in BER. The increase in BER shifts the average up slightly from the expected value and results in a large standard deviation. Furthermore, the logarithmic scaling amplifies these outliers, appearing more significant. Finally, it was concluded in the laser linewidth analysis, that with lower linewidth, results do not improve for an attacker when QEPS is present, however, when no encryption is applied, the results obtained by an eavesdropper will be superior. As expected, with a better-quality signal, where the linewidth is improved, an attacker is able to obtain a lower BER after tapping. However, with the addition of QEPS to the system, the phase space of the coherent state is scrambled, preventing the eavesdropper from obtaining any significant information even if the quality of their tap is improved. Thus, it can be concluded that linewidth and QEPS are independent of each other; if the eavesdropper has knowledge of the QEPS applied, their results will be better at lower linewidth, but if they have no knowledge of the QEPS applied, their results will have no difference at any linewidth. Based on these results and summarized in TABLE II, by selecting a smaller PM period value and a larger phase deviation value, the security can be maintained. Again, these results demonstrate the security of applying a quantum encryption to prevent an eavesdropper from obtaining information sent over the optical network using different parameters.

Besides the theoretical verified security, the practical security of QEPS must be considered. An eavesdropper has one main challenge in obtaining the secret information sent from Alice to Bob due to the phase randomization pattern. Eve can attempt to decrypt the encoded data through their own PM or through digital decryption, however, without knowledge of the pre-shared pQRNG seed, Eve will be unable to obtain the correct bits. Although slightly simpler for QEPS, where the attacker can also decode and compensate digitally, the obstacle of having the correct random seed remains. With a cryptographic-secure pQRNG [21], the sufficiently large entropy will deter attackers from attempting to decrypt the encryption even when part of the initial or running state becomes available due to future states being unpredictable. The said pQRNG is capable to take a secret of up to 16 kB long, perfect for QEPS. Moreover, the complexity of this system can quickly be increased through randomizing the phase deviation and period parameter during operation. This additional randomization must be driven by a deterministic RNG component so that encryption and decryption can be seamless. It is recommended that the same RNG unit used to drive the PM should be used to randomize the phase deviation and period parameter in order to reduce resources. Furthermore, tested in both simulation and experiment previously, if there is even one symbol shifted in the phase de-randomization, then a maximum BER of 0.5 is obtained. This is similar to [26] which discusses how synchronization is a monumental challenge for an eavesdropper in decoding.

Finally, we will briefly discuss other forms of physical layer attacks on the system. Common vulnerabilities in optical networks that can still be exploited such as gain competition in erbium-doped fiber amplifiers, interchannel crosstalk, correlated jamming and denial of service through fiber damage [28]. All of these disruptions can cause major issues to the optical network service; however, the network security and robustness can be maintained through traditional network routing algorithms [29], minimizing system disruptions. QEPS can be integrated into current infrastructure allowing it to leverage these technologies. With the vast array of interconnected optical fiber, service disruptions can be quickly rerouted to reduce physical impairment. With the addition of QEPS to current infrastructure, there is little to no added complexity with the advantage of increased security at the optical layer. Only one optical component is added to the system for the encoding and decoding can be performed in DSP. Thus, QEPS can be used as a physical layer security against attacks where the malicious party target the data in the network, however, for attacks which are targeted to disrupt a network, current solutions can be still leveraged.

## VI. Summary

We have proposed a theoretical model for a new optical encryption layer scheme over existing coherent optical communication channel utilizing randomized phase encoding. QEPS's quantum encryption scheme is compatible with common modulation formats such as PSK and QAM. Both PSK and QAM formats were tested at 28 GBaud. A common eavesdropping scenario was considered and it was demonstrated that the system was secure against an eavesdropper with no knowledge of the randomization seed. Similar performance trends were found for QEPS as CTF-QKD involving the period and phase deviation. With larger phase deviation and smaller periods, the security of the system increases with a trade-off of higher BER. In contrast to CTF-QKD, it was concluded in the CW power analysis that a minimum power was required for operation instead of a window of operation range. Constellation diagrams were also compared to demonstrate the effect of quantum encryption has only the optical network. Preliminary experimental validation was performed, and results were in line with simulation results. Other physical layer attacks were also discussed where the addition of QEPS does not add vulnerability to currently used conventional coherent optical communication networks, however, QEPS can still be affected by common disruptions. QEPS provides a unique solution to telco operators to control the data security of their system over the infrastructure layer. This model will be employed experimentally to validate our numerical results and we are in the process of experimental exploration where the results will be reported separately. Future work will also explore the improvement of security for QAM data modulations.

## VII. Acknowledgements

The authors thank Ciena for providing the WaveLogic modem used in the experiments. The authors thank the Mitacs Accelerate Program for the financial support for the project.

## References

[1] K. Holl, *OSI Defense in Depth to Increase Application Security,* Singapore: SANS Institute, 2003.

[2] K. Fouli and M. Maier, "OCDMA and Optical Coding: Principles, Applications, and Challenges," *IEEE Communications Magazine,* vol. 45, no. 8, pp. 27-34, 2007.

[3] A. Argyris, D. Syvridis, L. Larger, V. Annovazzi-Lodi, P. Colet, I. Fischer, J. García-Ojalvo, C. R. Mirasso, L. Pesguera and K. A. Shore, "Chaos-based communications at high bit rates using commercial fibre-optic links," *Nature,* vol. 438, pp. 343-346, 2005.

[4] B. Wu, Z. Wang, Y. Tian, M. P. Fok, B. j. Shastri, D. R. Kanoff and P. R. Prucnal, "Optical steganography based on amplified spontaneous emission noise," *Optics Express,* vol. 21, no. 2, pp. 2065-2071, 2013.

[5] Z. Li and G. Li, "Ultrahigh-speed reconfigurable logic gates based on four-wave mixing in a semiconductor optical amplifier," *IEEE Photonics Technology Letters,* vol. 18, no. 12, pp. 1341-1343, 2006.

[6] T. Shake, "Security performance of optical CDMA Against eavesdropping," *Journal of LIghtwave Technology,* vol. 23, no. 2, pp. 655-670, 2005.

[7] C. H. Bennett and G. Brassard, "Quantum Cryptography: Public Key Distribution and Coin Tossing," in *International Conference on Computers, Systems & Signal Processing*, Bangalore, India, 1984.

[8] I. B. Djordjevic, Physical-Layer Security and Quantum Key Distribution, Cham, Switzerland: Springer, 2019.

[9] H. Wang, Y. Pi, W. Huang, Y. Li, Y. Shao, J. Yang, J. Liu, C. Zhang, Y. Zhang and B. Xu, "High-speed Gaussian-modulated continuous-variable quantum key distribution with a local local oscillator based on pilot-tone-assisted phase compensation," *Optics Express,* vol. 28, no. 22, pp. 32882-32893, 2020.

[10] J. F. Dynes, W. W.-S. Tam, A. Plews, B. Fröhlich, A. W. Sharpe, M. Lucamarini, Z. Yuan, C. Radig, A. Straw, T. Edwards and A. J. Shields, "Ultra-high bandwidth quantum secured data transmission," *Scientific Reports,* vol. 6, no. 35149, pp. 1-6, 2016.

[11] M. Lucamarini, Z. L. Yuan, J. F. Dynes and A. J. Shields, "Overcoming the rate-distance limit of quantum key distribution without quantum repeaters," *Nature,* vol. 557, pp. 400-403, 2018.

[12] S. Pirandola, R. Laurenza, C. Ottaviani and L. Banchi, "Fundamental limits of repeaterless quantum communications," *Nature Communications,* vol. 8, no. 15043, 2017.

[13] S. Wang, Z.-Q. Yin, D.-Y. He, W. Chen, R.-Q. Wang, P. Ye, Y. Zhou, G.-J. Fan-Yuan, F.-X. Wang, W. Chen, Y.-G. Zhu, P. V. Morozov, A. V. Divochiy, Z. Zhou, G.-C. Guo and Z.-F. Han, "Twin-field quantum key distribution over 830-km fibre," *Nature Photonics,* vol. 16, pp. 154-161, 2022.

[14] "Quantum Key Distribution (QKD) and Quantum Cryptography (QC)," National Security Agency/Central Security Service, [Online]. Available: https://www.nsa.gov/Cybersecurity/Quantum-Key-Distribution-QKD-and-Quantum-Cryptography-QC/. [Accessed 23 March 2022].

[15] V. Scarani and C. Kurtsiefer, "The black paper of quantum cryptography: Real implementation problems," *Theoretical Computer Science,* vol. 560, pp. 27-32, 2014.

[16] S. Shi and N. Xiao, "10-Gb/s data transmission using optical physical layer encryption and quantum key distribution," *Optics Communications,* vol. 507, p. 127603, 2022.

[17] R. Kuang and N. Bettenburg, "Quantum Public Key Distribution using Randomized Glauber States," *2020 IEEE International Conference on Quantum Computing and Engineering (QCE)*, Denver, CO, USA, 2020, pp. 191-196, doi: 10.1109/QCE49297.2020.00032.

[18] A. Chan, M. Khalil, K. A. Shahriar, L. R. Chen, D. V. Plant and R. Kuang, "Security Analysis of a Next Generation TF-QKD for Secure Public Key Distribution with Coherent Detection over Classical Optical Fiber Network," in *2021 7th ICCC*, Chengdu, 2021.

[19] M. Khalil, A. Chan, K. A. Shahriar, L. R. Chen, D. V. Plant and R. Kuang, "Security Performance of Public Key Distribution in Coherent Optical Communications Links," in *2021 3rd ICCCI*, Nagoya, Japan, 2021.

[20] A. Zhao, N. Jiang, S. Liu, Y. Zhang and K. Qiu, "Physical-Layer Secure Optical Communication Based on Private Chaotic Phase Scrambling," in *26th Optoelectronics and Communications Conference*, Hong Kong, 2021.

[21] R. Kuang, D. Lou, A. He, C. McKenzie and M. Redding, "Pseudo Quantum Random Number Generator with Quantum Permutation Pad," *2021 IEEE International Conference on Quantum Computing and Engineering (QCE)*, Broomfield, CO, USA, 2021, pp. 359-364, doi: 10.1109/QCE52317.2021.00053.

[22] B. Goebel, R.-J. Essiambre, G. Kramer, P. J. Winzer and N. Hanik, "Calculation of Mutual Information for Partially Coherent Gaussian Channels With Applications to Fiber Optics," *IEEE Transactions on Information Theory,* vol. 57, no. 9, pp. 5720-5736, 2011.

[23] G. Colavolpe and R. Raheli, "The capacity of noncoherent channels," *European Transactions on Telecommunications,* vol. 12, no. 4, pp. 289-296, 2001.

[24] N. Ferguson and B. Schneier, Practical Cryptography, Indianapolis: Wiley, 2003.

[25] K. A. Shahriar, M. Khalil, A. Chan, L. R. Chen, R. Kuang and D. V. Plant, "Physical-Layer Secure Optical Communication Based on Randomized Phase Space in Pseudo-3-Party Infrastructure," in *Conference on Lasers and Electro-Optics*, San Jose, 2022.

[26] A. Chan, M. Khalil, K. A. Shahriar, L. R. Chen, D. V. Plant and R. Kuang, "On the Security of an Optical Layer Encryption using Coherent-based TF-QKD in Classical Optical Fiber Links," in *2022 4th ICCCI*, Chiba, Japan, 2022.

[27] S. Camatel and V. Ferrero, "Narrow Linewidth CW Laser Phase Noise Characterization Methods for Coherent Transmission System Applications," *Journal of Lightwave Technology,* vol. 26, no. 17, pp. 3048-3055, 2008.

[28] N. Skorin-Kapov, J. Chen and L. Wosinska, "A New Approach to Optical Network Security: Attack-Aware Routing and Wavelength Assignment," *IEEE/ACM Transactions on Networking,* vol. 18, no. 3, pp. 750-760, 2010.

[29] D. Medhi and K. Ramasamy, Network Routing: Algorithms, Protocols, and Architectures, Cambridge: Elsevier, 2018.